\documentclass[aps,pre,showpacs,twocolumn,superscriptaddress,
              floats,floatfix]{revtex4}
\usepackage{amssymb,amsmath}
\usepackage{graphics}
\usepackage[usenames,dvipsnames]{color}
\usepackage{subfigure}
\usepackage{epsfig,ulem}
\usepackage{hyperref}
\usepackage{breakurl}

\begin{document}
\title{Two-dimensional, homogeneous, isotropic fluid turbulence with polymer additives}
\author{Anupam Gupta}
\email{agupta@roma2.infn.it}
\affiliation{Centre for Condensed Matter Theory, Department of Physics, Indian
Institute of Science, Bangalore 560012, India.} 
\affiliation{
Department of Physics,
University of Rome``Tor Vergata''
Via della Ricerca Scientifica 1,
00133 Roma Italy
}
\author{Prasad Perlekar} 
\email{perlekar@tifrh.res.in}
\affiliation{TIFR Centre for Interdisciplinary Sciences, 21 Brundavan Colony,
 Narsingi, Hyderabad 500075, India}
\author{Rahul Pandit}
\email{rahul@physics.iisc.ernet.in.}
\affiliation{Centre for Condensed Matter Theory, Department of Physics, Indian
Institute of Science, Bangalore 560012, India.} 
\affiliation{Jawaharlal Nehru Centre For Advanced Scientific Research, Jakkur, Bangalore, India.}
\begin{abstract}
We carry out the most extensive and high-resolution direct numerical 
simulation, attempted so far, of homogeneous, isotropic turbulence in 
two-dimensional fluid films with air-drag-induced friction and with 
polymer additives. Our study reveals that the polymers (a) reduce 
the total fluid energy, enstrophy, and palinstrophy, (b) modify the
fluid energy spectrum both in inverse- and forward-cascade r\'egimes, (c)
reduce small-scale intermittency, (d) suppress regions of large vorticity 
and strain rate, and (e) stretch in strain-dominated regions. 
We compare our results with earlier experimental studies and propose new 
experiments.
\end{abstract}
\maketitle

\section{Introduction}
\label{sec:Introduction}

Polymer additives have remarkable effects on turbulent flows: in wall-bounded
flows they lead to drag reduction~\cite{toms,virk75}; in homogeneous, isotropic
turbulence they give rise to dissipation reduction, a modification of the
energy spectrum, and a suppression of small-scale
structures~\cite{hoyt,doorn99,kalelkar05,perlekar06,perlekar10,zhang10,boffetta12,bodengroup,procacciapoly,benzi14,watanable14}.
These effects have been studied principally in three-dimensional (3D) flows;
their two-dimensional (2D) analogs have been studied only over the past decade
in experiments~\cite{amarouchene02,kellayexpt,jun06} on and direct numerical
simulations (DNSs) ~\cite{mussachiothesis,berti08,boffetta0305,kellaydns} of
fluid films with polymer additives. It is important to investigate the
differences between 2D and 3D fluid turbulence with polymers because the
statistical properties of fluid turbulence in 2D and 3D are qualitatively
different~\cite{boffettaecke}: the inviscid, unforced 2D Navier-Stokes (NS)
equation admits more conserved quantities than its 3D counterpart; one
consequence of this is that, from the forcing scales, there is a flow of energy
towards large length scales (an inverse cascade) and that of enstrophy towards
small scales (a forward cascade). We have, therefore, carried out the most
extensive and high-resolution DNS study of homogeneous, isotropic turbulence in
the incompressible, 2D NS equation with air-drag-induced friction and polymer
additives, described by the finitely-extensible-nonlinear-elastic-Peterlin
(FENE-P) model for the polymer-conformation tensor. We find that the
inverse-cascade part of the energy spectrum in 2D fluid turbulence is
suppressed by the addition of polymers.  We show, for the first time, that the
effect of polymers on the forward-cascade part of the fluid energy spectrum in
2D is (a) a slight reduction at intermediate wave numbers and (b) a significant
enhancement in the large-wave-number range, as in 3D; the high resolution of
our simulation is essential for resolving these spectral features
unambiguously.  In addition, we find dissipation-reduction-type
phenomena~\cite{perlekar06,perlekar10}: polymers reduce the total fluid energy
and energy- and mean-square-vorticity- or enstrophy-dissipation rates, suppress
small-scale intermittency, and decrease high-intensity vortical and
strain-dominated r\'egimes. Our probability distribution functions (PDFs) for
$\sigma^2$ and $\omega^2$, the squares of the strain rate and the vorticity,
respectively, agree qualitatively with those in experiments~\cite{jun06}. We
also present PDFs of the Okubo-Weiss parameter $\Lambda = (\omega^2 -
\sigma^2)/8$, whose sign determines whether the flow in a given region is
vortical or strain-dominated~\cite{perlekar09,okubo}, and PDFs of the polymer
extension; and we show explicitly that polymers stretch preferentially in
strain-dominated regions.  

The remaining part of this paper is organized as follows. In
Sec.~\ref{sec:Model} we define the equations we use for polymer additives in a
fluid and we describe the numerical methods we use to study these equations.
Section~\ref{sec:Results} is devoted to the results of our study and
Sec.~\ref{sec:Conclusions} contains a discussion of our principal results. 

\section{Model and Numerical Methods}
\label{sec:Model}

\begin{table*}
{\setlength{\tabcolsep}{0.33em}
   \begin{tabular}{c c c c c c c c c c c c c}
    \hline
    $ $ &$N$ & $L$ & $\tau_P$ & $\delta{t} \times 10^4$ & $E_{inj}$ & $\nu\times10^{4}$ & ${\mathcal Wi}$
& $c$ & $Re_{\lambda}$ & $ k_{max}\eta_d$\\
   \hline \hline
    {\tt R1}  & $512$ & $6$ & $2$ & $10.0$ & $0.008$ &  $10.0$	&	$4.71$ & $~0.1$ & $107,~85$
& $3.4,~3.6$\\
    {\tt R2}  & $1024$ & $100$ & $1,2,4$ & $1.0$ & $0.005$ &  $5.0$	& $2.26~4.52~9.04~~$ &
$~0.1$ & $221,~121,~53~,~~38~~$ & $5.1,5.3,5.4,5.5$\\
    {\tt R3}  & $2048$ & $100$ & $1$ & $1.0$ & $0.003$ &  $5.0$	& $1.81$ & $~0.4$ & $147,~60$
&$14.1,~14.8 $\\
    {\tt R4}  & $2048$ & $100$ & $1$ & $1.0$ & $0.0015$ &  $5.0$	& $1.35$ & $~0.2$ & $86~,~~54$
& $13.2,~13.6$\\
    {\tt R5}  & $4096$ & $100$ & $1$ & $1.0$ & $0.005$ &  $5.0$	&	$2.21$ & $~0.2$ & $233,~91$ &
$20.2,~20.9$\\
    {\tt R6}  & $4096$ & $100$ & $1$ & $1.0$ & $0.002$ &  $5.0$	&	$1.53$ & $~0.2~0.4$ &
$108,~62,~45$ & $24.8,~25.8,~ 26.1$\\
    {\tt R7}  & $4096$ & $10$ & $1$ & $1.0$ & $0.002$ &  $5.0$	&	$1.53$ & $~0.4$ &
$108,~~90$ & $24.8,~ 26.2$\\
    {\tt R8}  & $4096$ & $100$  & $1$ & $0.5$ & $0.005$ &  $1.0$	&	$2.91$ &
$~0.1~0.4$ & $1451,~1367,~1311$
& $8.0,~8.3,~8.5$\\
    {\tt R9}  & $4096$ & $10$  & $1$ & $0.5$ & $0.005$ &  $1.0$	&	$2.91$ & $~0.1$ & $1451,~1407$
& $8.0,~8.2$\\
    {\tt R10}  & $16384$ & $100$ & $1$ & $0.5$ & $0.002$ &  $5.0$	&	$1.56$ & $~0.2$ & $106,~61$
& $96.4,~102.7$\\
\hline
\end{tabular}}
\caption{\small
Parameters for our DNS runs {\tt R1-R10} with the friction coefficient $\alpha = 0.01$.
$N^2$ is the number of collocation points, $\delta t$ the time step, $E_{inj}$ the
energy-injection rate, $\nu$ the kinematic viscosity, and $c$ the concentration parameter.
The Taylor-microscale Reynolds number is $Re_{\lambda}\equiv u_{rms} \lambda/\nu$, where
$\lambda = ({\int E(k)dk}/{\int k^2 E(k)dk})^{1/2}$ and the Weissenberg number is ${\mathcal
Wi} \equiv \tau_P \sqrt{\epsilon^f/\nu}$, where $\epsilon^f$ is the energy dissipation rate
per unit mass for the fluid. The dissipation scale is $\eta_d \equiv
(\nu^3/\epsilon)^{1/4}$ and $k_{max} = N/3$. 
}
\label{table:para}
\end{table*} 

The 2D incompressible NS and FENE-P equations can be written in terms of the
stream-function $\psi$ and the vorticity ${\boldsymbol{\omega}} = \nabla
\times {\bf u}({\bf x},t)$, where ${\bf u}\equiv(-\partial_y \psi, \partial_x
\psi)$ is the fluid velocity at the point ${\bf x}$ and time $t$, as follows:
\begin{eqnarray}
D_t{\bf \omega} &=& \nu \nabla^2 {\bf \omega}+
              \frac{\mu}{\tau_P} \nabla \times \nabla.[f(r_P){\cal C}] 
	      - \alpha \omega   + F_{\omega} ;    
						    \label{ns}\\
\nabla^2 {\bf \psi}  &=& {\bf \omega}; 
						\label{poisson}\\
D_t{\cal C}&=& {\cal C}. (\nabla {\bf u}) + 
                {(\nabla {\bf u})^T}.{\cal C} - 
                \frac{{f(r_P){\cal C} }- {\cal I}}{\tau_P}.
                                                   \label{FENE}
\end{eqnarray}
Here $D_t\equiv\partial_t + {\bf u}.\nabla$, the uniform solvent density
$\rho = 1$, $\alpha$ is the coefficient of friction, $\nu$ the kinematic
viscosity of the fluid, $\mu$ the viscosity parameter for the solute
(FENE-P), and $\tau_P$ the polymer relaxation time; to mimic 
experiments~\cite{jun06}, we use a Kolmogorov-type
forcing $F_{\omega}\equiv k_{inj} F_0 \cos(k_{inj}y)$, with amplitude $F_0$;
the energy-injection wave vector is $k_{inj}$ (the length scale
$l_{inj}\equiv 2\pi /k_{inj}$); the superscript $T$ denotes a transpose,
${\cal C}_{\beta\gamma}\equiv {\langle{R_\beta}{R_\gamma}\rangle}$ are the
elements of the polymer-conformation tensor (angular brackets indicate an
average over polymer configurations), $\cal I$ is the identity tensor, $f(r_P)\equiv{(L^2
-2)/(L^2 - r_P^2)}$ is the FENE-P potential, and $r_P \equiv \sqrt{{\rm Tr}(\cal C)}$ and
$ L $ are, respectively, the length 
and the maximal possible extension of the polymers; and
$c\equiv\mu/(\nu+\mu)$ is a dimensionless measure of the polymer
concentration~\cite{vai03} .

We use periodic boundary conditions, a
square simulation domain with side ${\mathbb L}=2\pi$ and $N^2$ collocation
points, a fourth-order, Runge-Kutta scheme, with time step $\delta t$,
for time marching, an explicit, fourth-order, central-finite-difference
scheme in space, and the Kurganov-Tadmor (KT) shock-capturing
scheme~\cite{kur00} for the advection term in Eq.~(\ref{FENE}); the KT scheme
(Eq. (7) of Ref.~\cite{perlekar10}) resolves sharp gradients in ${\cal
C}_{\beta\gamma}$ and thus minimizes dispersion errors, which increase with
$L$ and $\tau_P$. We solve Eq.~(\ref{poisson}) in Fourier space by using 
the FFTW library~\cite{fftw}. We choose $\delta t
\simeq 10^{-3} ~{\rm to}~ 5 \times 10^{-5}$ so that $r_P$ does not
become larger than $L$ (Table~\ref{table:para}).
We preserve the symmetric-positive-definite (SPD) nature of $\cal C$ by adapting to 2D the
Cholesky-decomposition scheme of Refs.~\cite{vai03,perlekar06,perlekar10}: We define
${\cal J} \equiv f(r_P) {\cal C}$, so Eq.~(\ref{FENE}) becomes
\begin{equation} 	
D_t{\cal J} = {\cal J}. (\nabla {\bf u}) 
+ ({\nabla \bf u})^T .{\cal J} -s({\cal J} - {\cal I})+ q {\cal J},
\label{conj} 
\end{equation} 
where $s=(L^2 -2+ j^2)/(\tau_P L^2)$, $q=[d/(L^2 -2)-(L^2 -2+ j^2)(j^2
-2)/(\tau_P L^2(L^2 -2))]$, $j^2\equiv Tr({\cal J})$, and $d = Tr[ {\cal J}.
(\nabla{\bf u}) + (\nabla{\bf u})^T .{\cal J}].$ Given that ${\cal C}$ and
hence ${\cal J}$ are SPD matrices, we can write ${\cal J}= {\cal LL}^T$,
where ${\cal L}$ is a lower-triangular matrix with elements $\ell_{ij}$, such
that  $\ell_{ij}=0$ for $j>i$; Eq.\eqref{conj} now yields $(1\le i \le2$ and
$ \Gamma_{ij}\equiv \partial_i u_j )$  
\begin{eqnarray}
\nonumber
{D_t \ell_{11}}  &=& \Gamma_{11} \ell_{11} + \Gamma_{21} \ell_{21}
+ \frac{1}{2}\Big[(q-s)\ell_{11}+ \frac{s}{\ell_{11}}\Big], \\
\nonumber    
{D_t \ell_{21}}  &=& \Gamma_{12} \ell_{11} + 
\Gamma_{21} \frac {\ell_{22}^2}{\ell_{11}} + 
\Gamma_{22} \ell_{21}\\
\nonumber
&&{}+ \frac{1}{2}\Big[(q-s)\ell_{21} - s \frac{\ell_{21}}{\ell^2_{11}} \Big], \\    
\nonumber    
{ D_t \ell_{22}}  &=& - \Gamma_{21} \frac {\ell_{21} \ell_{22}}{\ell_{11}} 
+ \Gamma_{22} \ell_{22} \\
&&{}+ \frac{1}{2}\Big[(q-s)\ell_{22} - \frac {s}{\ell_{22}}  \left( 1 + \frac {\ell^2_{21}}{\ell^2_{11}} \right) 
\Big].
\label{ellij}
\end{eqnarray}
Equation\eqref{ellij} preserves the SPD nature of  $\cal C$ if $\ell_{ii} >
0$, which we enforce~\cite{perlekar06,perlekar10} by considering the
evolution of $\ln(\ell_{ii})$ instead of $\ell_{ii}$.

\begin{figure*}
\hspace{-0.7cm}
\includegraphics[width=0.48\linewidth]{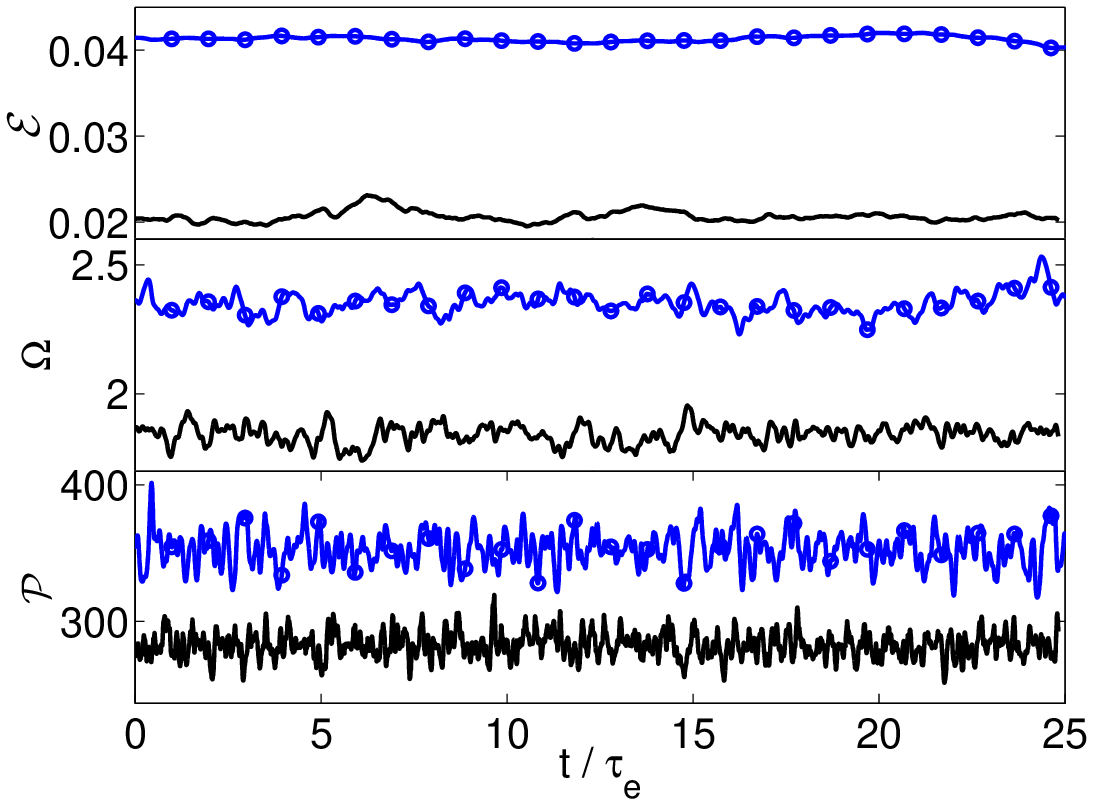}
\put(-45,150){ { {\large (a)} } }
\includegraphics[width=0.48\linewidth]{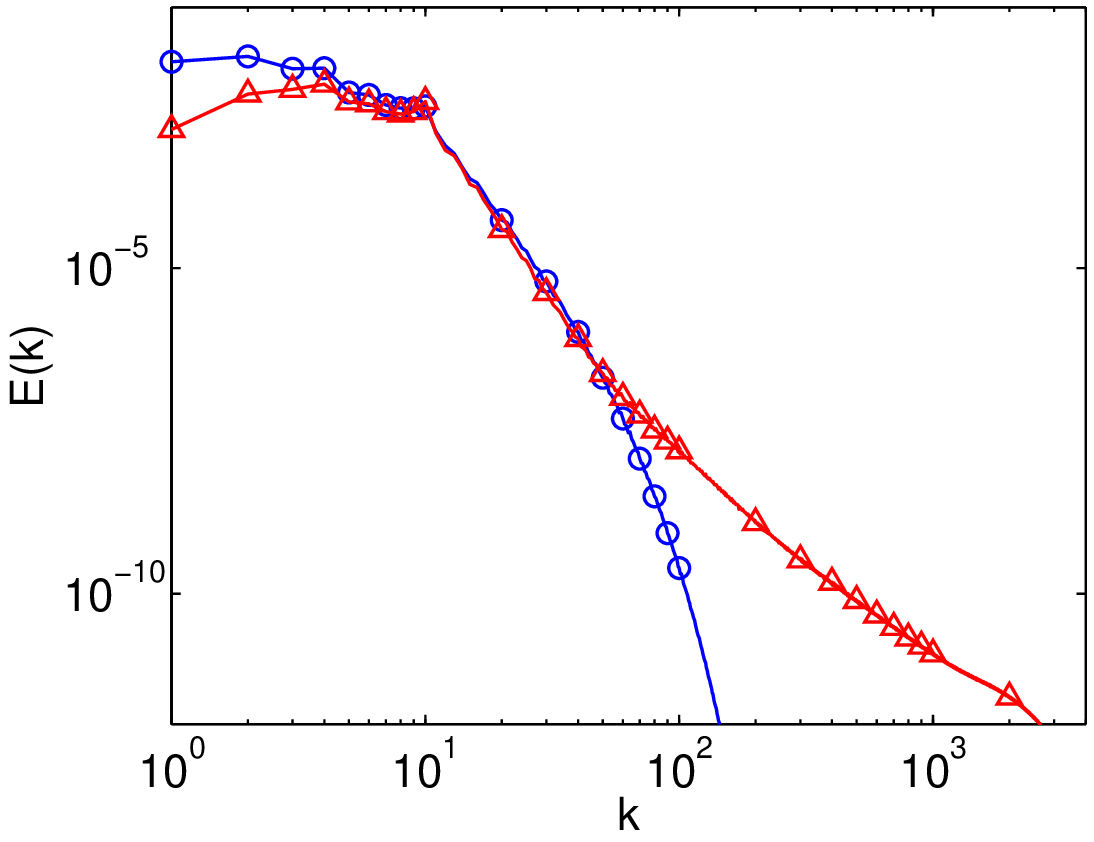}
\put(-45,160){ { {\large (b)} } }\\
\includegraphics[width=0.48\linewidth]{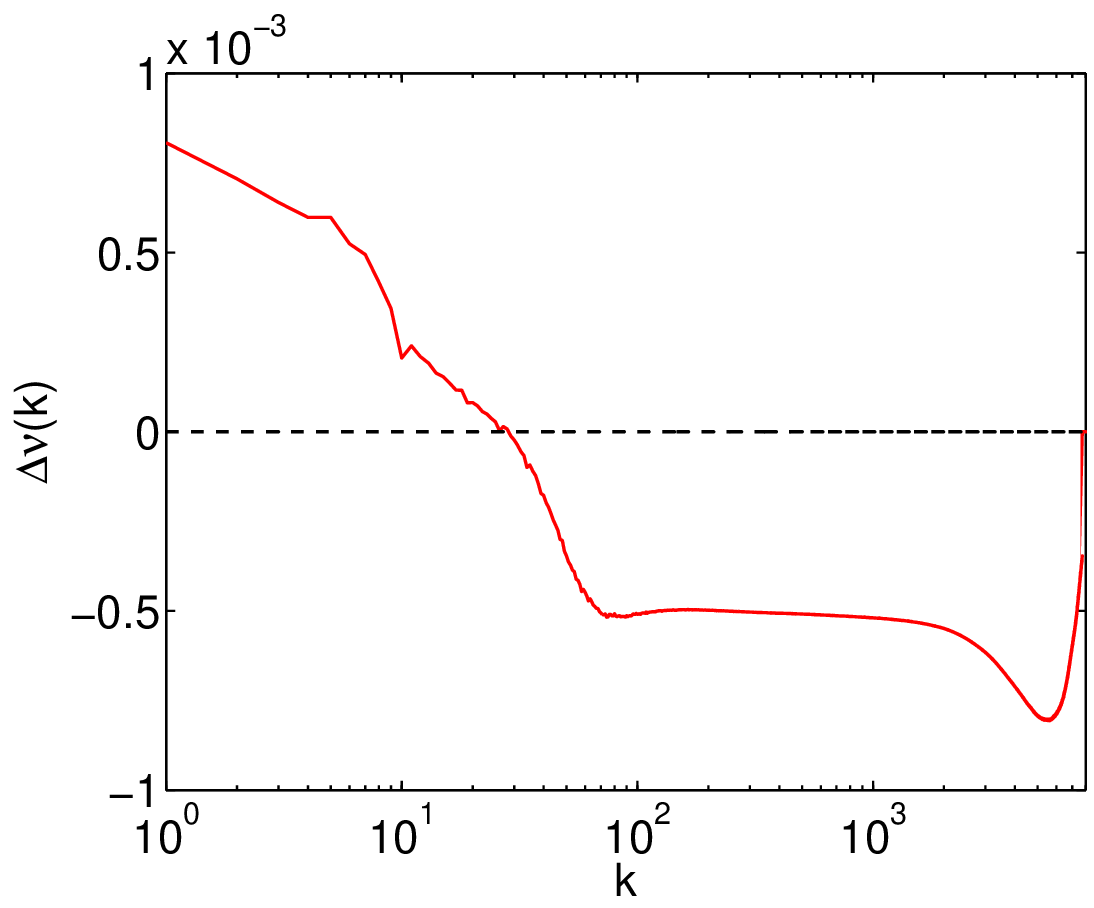}
\put(-45,160){ { {\large (c)} } }
\includegraphics[width=0.48\linewidth]{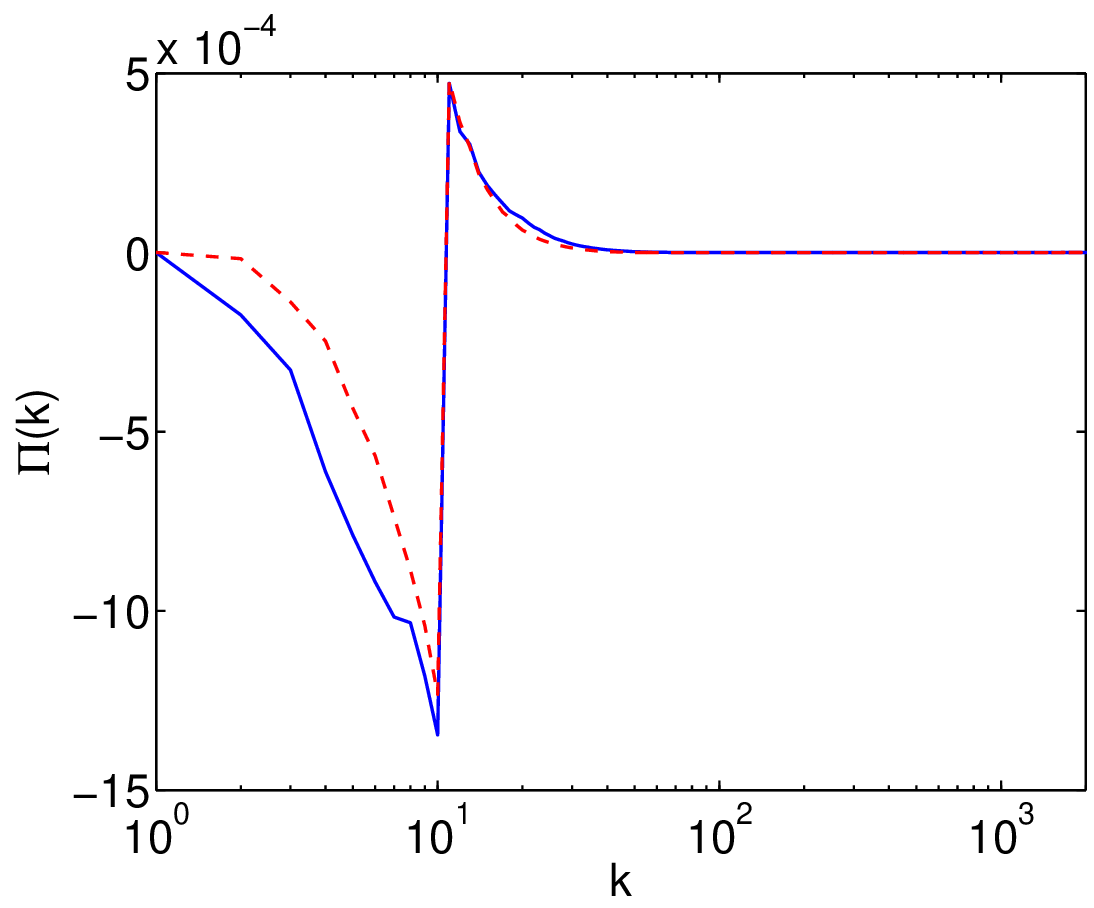}
\put(-45,160){ { {\large (d)} } }
\caption{\label{fig:eddr}\small(Color online) (a) Plots versus 
time $t/ \tau_e$ of the total kinetic energy $\mathcal E$ of
the fluid (top panel), the enstrophy $\Omega$ (middle panel), and the
palinstrophy $\mathcal P$ (bottom panel) for $c=0$ (upper
curve, blue circles for  run ${\tt R7}$) and $c=0.4$ (lower
curve, black line for run ${\tt R7}$); (b) log-log (base 10) plots of the
energy spectra $E(k)$ versus $k$ for $c = 0.2$ (red triangles for run {\tt
R10}) and $c=0$ (blue circles for run {\tt R10}); 
(c) polymer contribution to the scale-dependent viscosity $\Delta \nu(k)$ versus $k$ for $c = 0.2$
(red line for run {\tt R10}), $\Delta \nu(k) = 0$ is shown as black dotted line; and 
(d) energy flux $\Pi(k)$ versus $k$ for $c = 0.2$ (red dotted line for run {\tt R10}) and $c=0$ (blue
line for run {\tt R10}).}
\end{figure*}
\begin{figure*}
\includegraphics[width=0.48\linewidth]{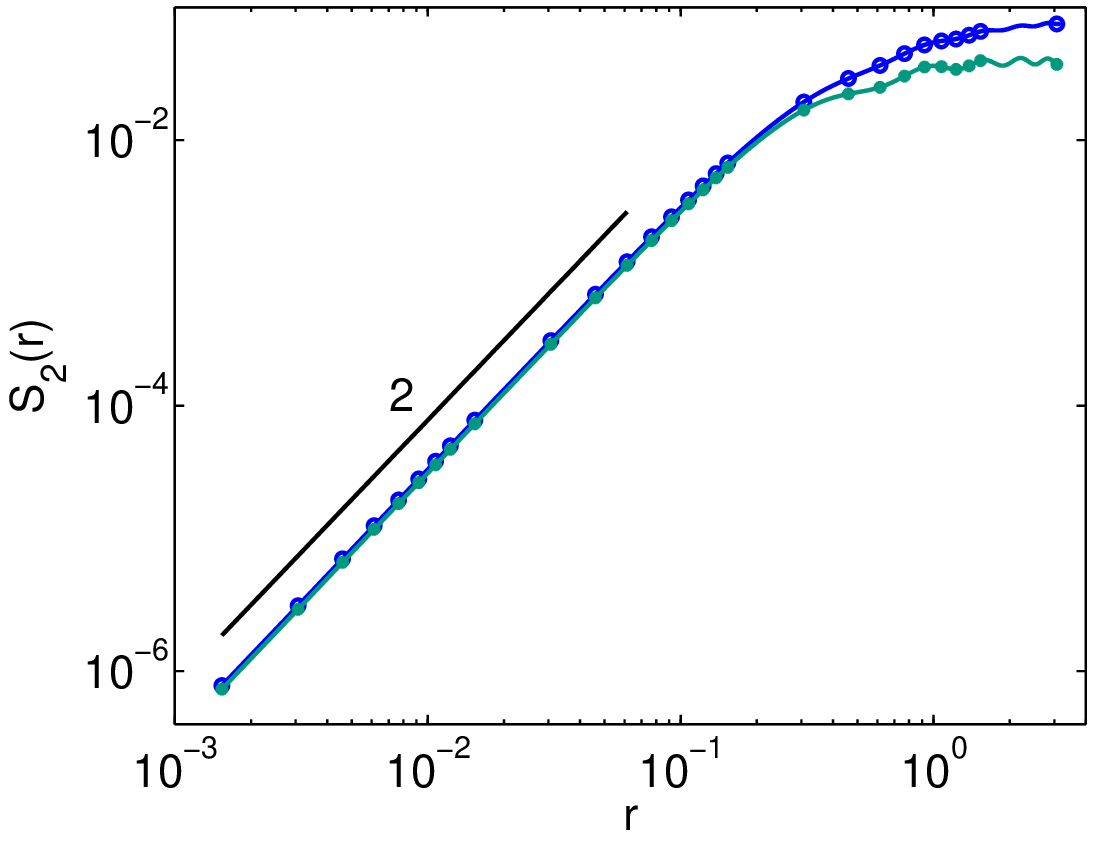}
\put(-195,160){ { {\large (a)} } }
\hspace{0.4cm}{}
\includegraphics[width=0.48\linewidth]{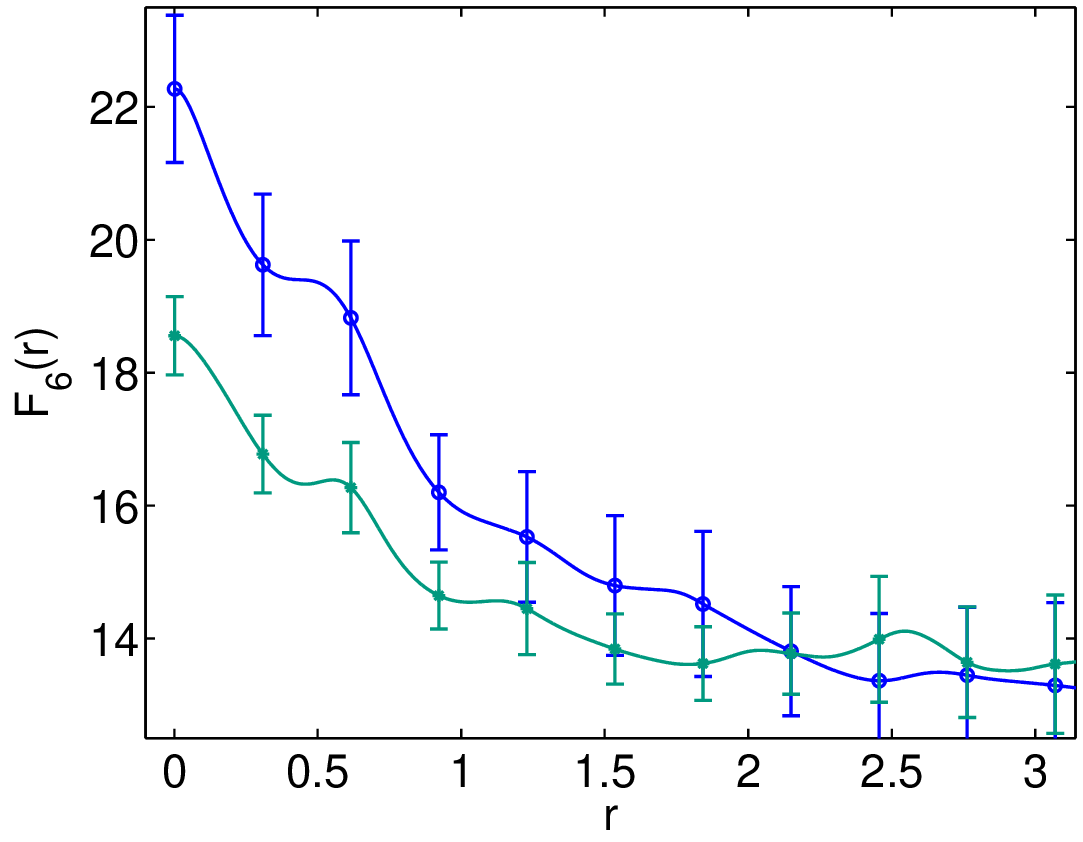}
\put(-45,160){ { {\large (b)} } }
\caption{\label{fig:s2fun}\small(Color online) (a) Plots of the second-order
velocity structure function $S_2(r)$ versus $r$ for $c=0$ (blue circle for run
{\tt R7}) and $c=0.2$ (green asterisks for run {\tt R7}); the line with slope
$2$ is shown for comparison; (b) plots of the hyeprflatness $F_6(r)$ versus $r$ for
$c=0$ (blue circles for run {\tt R7}) and $c=0.2$ (green asterisks for run {\tt
R7}).}
\end{figure*}

We have tested explicitly that the statistical properties we measure do not
depend on the resolutions we use for our DNS. We check this both by increasing
and decreasing this resolution.  Indeed, our DNS uses the highest resolution
that has been attempted so far for this problem (it uses 256 times as many
collocation points as those in Ref.~\cite{berti08}). Furthermore, the
Kurganov-Tadmor shock-capturing scheme that we use controls any dispersive
errors, because of sharp gradients in the polymer-conformation tensor, as in
similar three-dimensional studies~\cite{perlekar06,vai03}.

We maintain a constant energy-injection rate $E_{inj} \equiv \langle{\bf F_{u}}
\cdot \bf u\rangle$ with $F_{\omega} = \nabla \times {\bf F_u} $; the system
attains a nonequilibrium, statistically steady state after $\simeq
2\tau_e-3\tau_e$, where the box-size eddy-turnover time $\tau_e \equiv {\mathbb
L}/u_{rms}$ and $u_{rms}$ is the root-mean-square velocity.

In addition to ${\boldsymbol \omega}({\bf x},t)$, ${\bf \psi}({\bf x},t)$, and
${\cal C}({\bf x},t)$ we obtain ${\bf u (\bf x},t)$, the fluid-energy spectrum
$E(k)\equiv\sum_{k-1/2< k' \le k+1/2} k'^2 \langle |\hat {\bf \psi}({\bf
k'},t)|^2 \rangle _t$, where $\langle \rangle_t$ indicates a time average over
the statistically steady state, the total kinetic energy ~${\mathcal E}(t)
\equiv \langle \frac 1 2 |{\bf u(\bf x},t)|^2\rangle_{\bf x}$, enstrophy
~${\Omega}(t) \equiv \langle \frac {1}{2}|\boldsymbol \omega({\bf
x},t)|^2\rangle_{\bf x}$, and palinstrophy ~${\mathcal P}(t) \equiv \langle
\frac {1}{2} |{\nabla \times \boldsymbol \omega}({\bf x},t)|^2\rangle_{\bf x}$,
where $\langle \rangle_{\bf x}$ denotes a spatial average, the PDF of scaled
polymer extensions $P(r_P/L)$, the PDFs of $\omega^2$, $\sigma^2$, and $\Lambda
= (\omega^2 - \sigma^2)/8$, where $\sigma^2 \equiv \sum_{ij} \sigma_{ij}
\sigma_{ij}$, and $\sigma_{ij} \equiv \partial_i u_j + \partial_j u_i $, the
PDF of the Cartesian components of $\bf u$, and the  joint PDF of $\Lambda$ and
$r_P^2$. We obtain the isotropic part of order-$p$, structure function $S_p(r)$
from longitudinal velocity increments as described in Ref~\cite{perlekar09}. We
concentrate on $S_2(r)$ and the hyperflatness $F_6(r)\equiv S_6(r)/[S_2(r)^3]$;
the latter is a measure of the intermittency at the scale $r$.

\section{Results}
\label{sec:Results}

In Fig. (\ref{fig:eddr}a) we show how ${\mathcal E}(t)$ (top panel),
$\Omega(t)$ (middle panel), and ${\mathcal P}(t)$ (bottom panel) fluctuate
about their mean values $\langle {\mathcal E}(t)\rangle_t$, $\langle
\Omega(t)\rangle_t$, and $\langle {\mathcal P}(t)\rangle_t$ for $c=0$ (pure
fluid) and $c=0.4$. Clearly, $\langle {\mathcal E}(t)\rangle_t$, $\langle
\Omega(t)\rangle_t$, and $\langle {\mathcal P}(t)\rangle_t$ decrease as $c$
increases.  Thus, polymers increase the effective viscosity of the solution;
but this na\"ive conclusion has to be refined, as will be shown later, because the effective viscosity
depends on the length scale~\cite{procacciapoly,perlekar06,perlekar10}. 

In Fig. (\ref{fig:s2fun}a), we plot $S_2(r)$ versus $r$ for $c=0$ (blue circles
and run $\tt R7$) and $c=0.2$ (green asterisks and run $\tt R7$); the dashed
line, with slope 2, is shown to guide the eye; this slope agrees with the
$S_2(r) \sim r^2$ form that we expect, at small $r$, by Taylor expansion. At
large values of $r$, $S_2(r)$ deviates from this $r^2$ behavior, more so for
$c=0.2$ than for $c=0$, in accord with experiments ~\cite{jun06}. Plots of
$F_6(r)$ versus $r$ (Fig. (\ref{fig:s2fun}b)), for $c=0$ (blue circles) and
$c=0.2$ (green asterisks and run $\tt R8$), show that, on the addition of
polymers, small-scale intermittency decreases as $c$ increases.  

\begin{figure*}
\hspace{-0.6cm}
\includegraphics[width=0.48\linewidth]{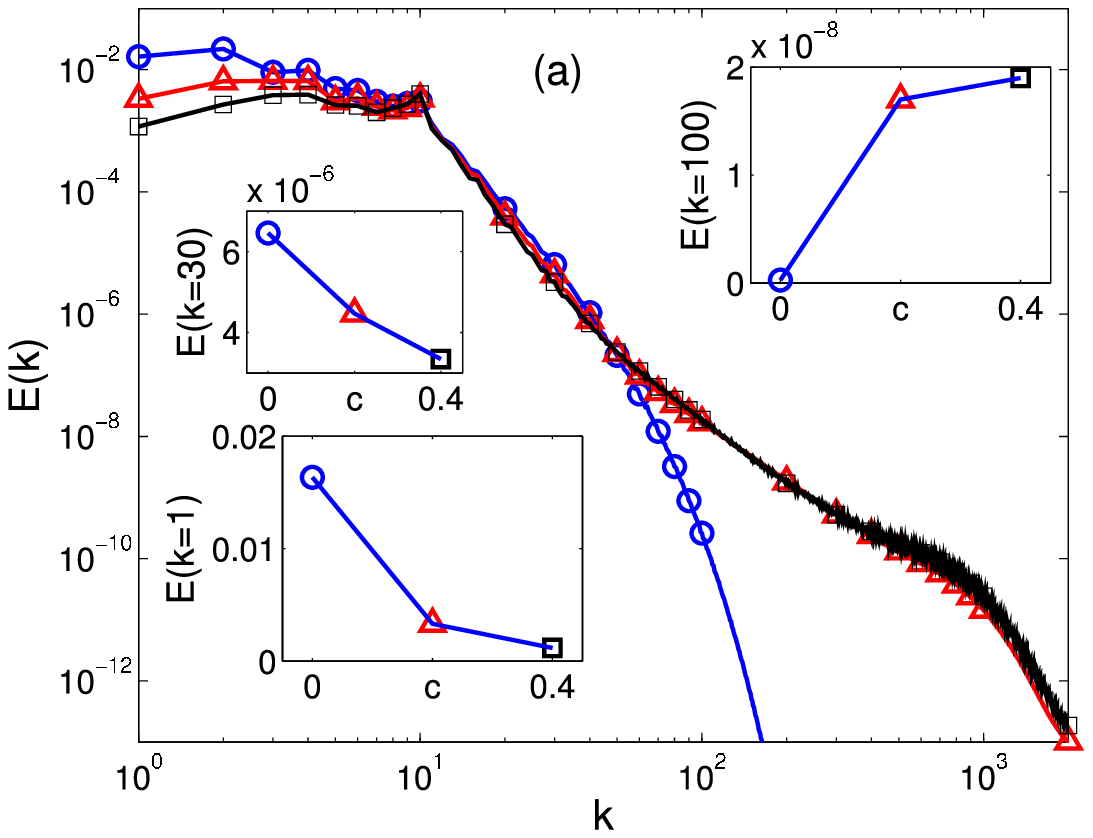}
\includegraphics[width=0.48\linewidth]{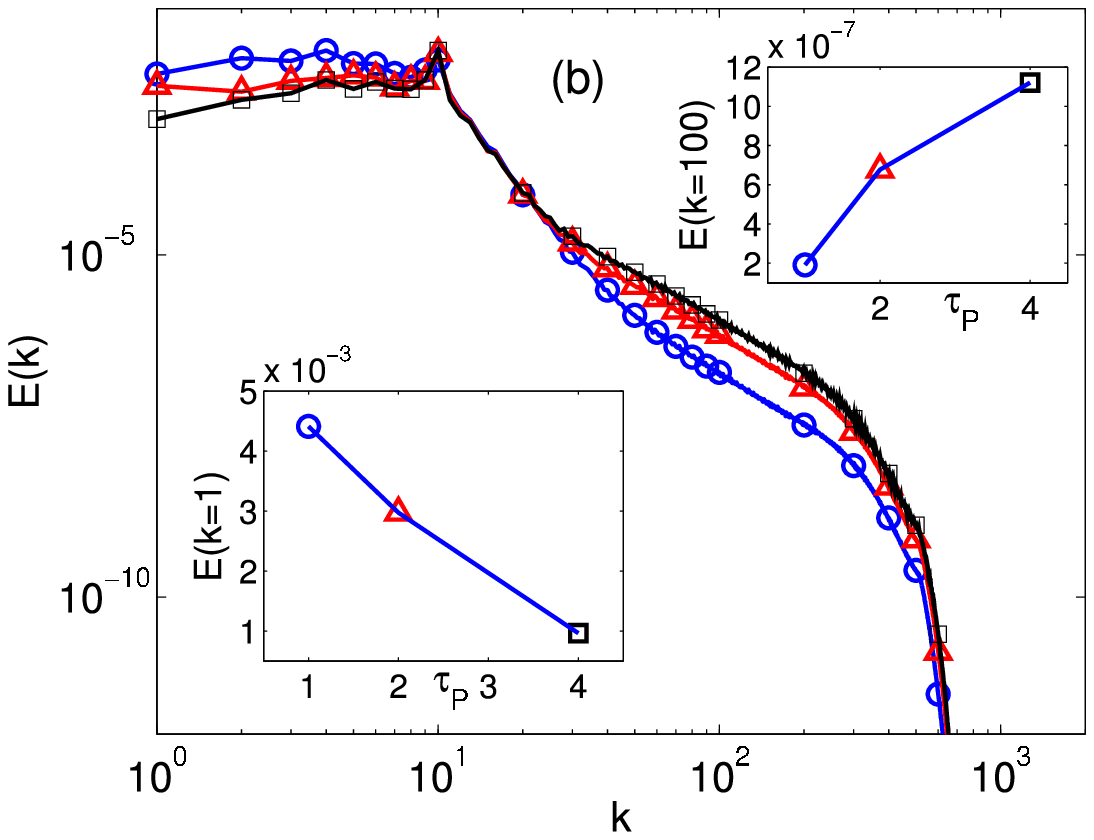}\\
\caption{\label{fig:spect}\small(Color online) 
(a) Log-log (base 10) plots of the energy spectra $E(k)$ versus $k$
for $c=0$  (blue circles for run ${\tt R6}$), $c = 0.2$ (red 
triangles for run ${\tt R6}$), and $c=0.4$ (black
squares for run ${\tt R6}$); plots of $E(k)$ versus $c$ for $\mathcal Wi = 1.53$ and $k =
1$ (left bottom inset), $\mathcal Wi = 1.53$ and $k = 30$ (left top inset), 
and $\mathcal Wi = 1.53$ and $k = 100$ (right top inset);
(b) log-log (base 10) plots of $E(k)$ versus $k$
for $\mathcal Wi = 2.26$ (blue circles for run $\tt R2$), $\mathcal Wi = 4.52$ (red 
triangles for run $\tt R2$), and $\mathcal Wi = 9.04$ (black squares for run $\tt R2$);
plots of $E(k)$ versus $\tau_P$ for $c = 0.4$ and $k = 1$ (left bottom inset) 
and $c = 0.4$ and $k = 100$ (right top inset).} 
\end{figure*}

In Fig. (\ref{fig:spect}a), we show how $E^p(k)$ changes, as we increase $c$ :
at low and intermediate values of $k$ (e.g., $k=1$ and $30$, respectively),
$E^p(k)$ decreases as $c$ increases; but, for large values of $k$ (e.g.,
$k=100$), it increases with $c$.  Figure (\ref{fig:spect}b) shows how $E^p(k)$
changes, as we increase $\tau_P$  with $c$ held fixed at $0.1$. At low values
of $k$ (e.g., $k=1$), $E^p(k)$ decreases as $\tau_P$ increases; but for large
values of $k$ (e.g., $k=100$) it increases with $\tau_P$.  
\begin{figure*}
\hspace{-0.6cm}
\includegraphics[width=0.48\linewidth]{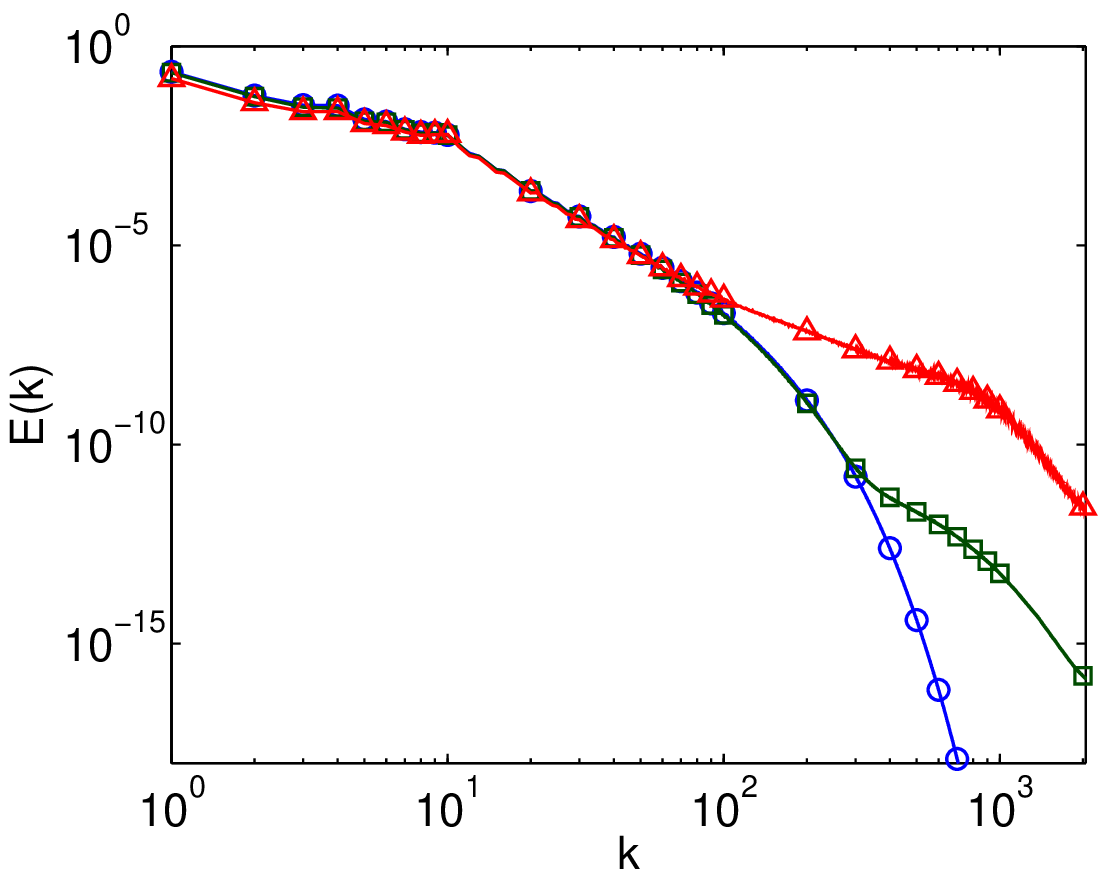}
\put(-45,160){ { {\large (a)} } }
\includegraphics[width=0.48\linewidth]{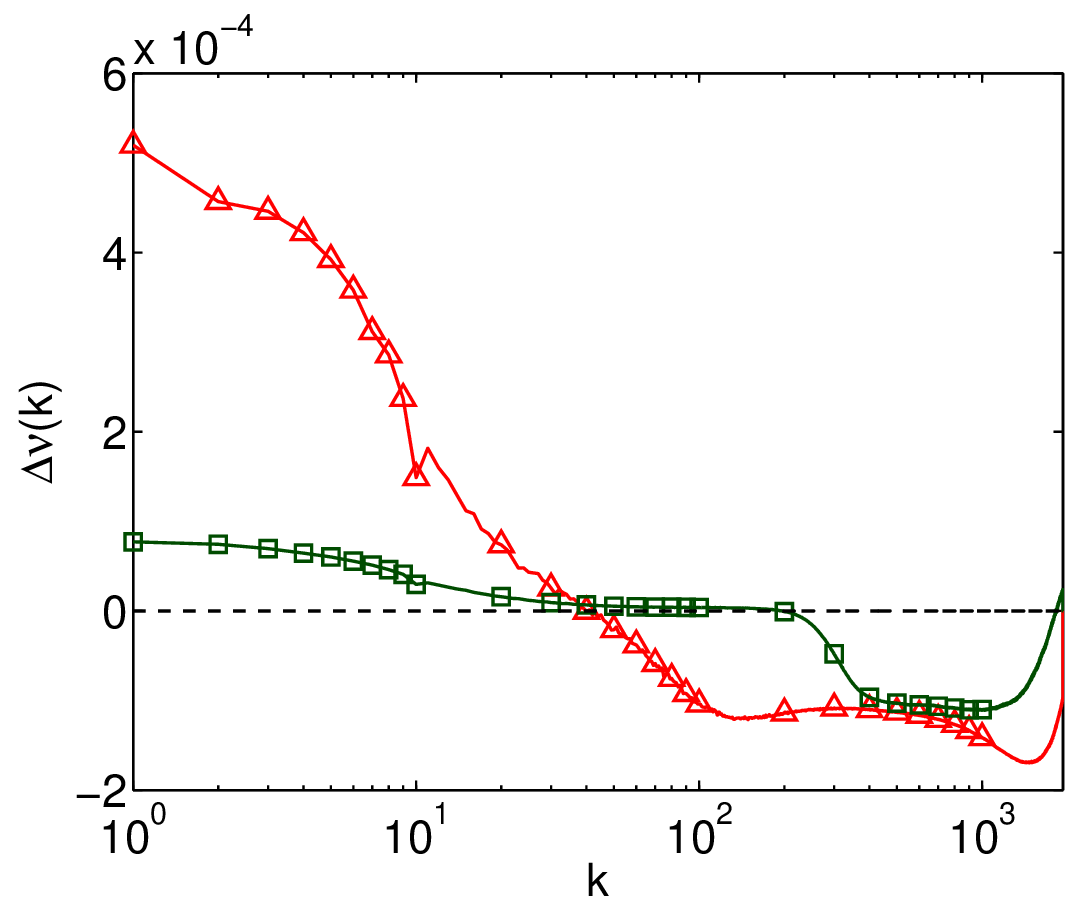}
\put(-45,160){ { {\large (b)} } }
\caption{\label{fig:spect_L2}\small(Color online) 
(a) Log-log (base 10) plots, for $c=0.2$ and $\mathcal Wi = 2.91$, of $E(k)$ versus $k$
for $L=100$ (red triangles for run $\tt R8$) and $L=10$ (green asterisks for run 
{\tt R9}); and $E(k)$ for $c=0$ (blue circles for run $\tt R8$); (b) plots, for $L=100$  (red triangles for run $\tt R8$) and $L=10$ (green asterisks for run {\tt R9}), of the scale-dependent correction to the viscosity $\Delta \nu(k)$ versus $k$.} 
\end{figure*}

In  Fig. (\ref{fig:spect_L2}a) we give plots, for $c=0.1$, of the spectra
$E^p(k)$ for $L=100$ (red triangles and run $\tt R8$) and $L=10$ (green
asterisks and run $\tt R9$); for comparison we also plot $E^f(k)$ for $c = 0$;
as $L$ increases, the difference between $E^p(k)$ and $E^f(k)$ increases at
large values of $k$. We see that the larger the value of 
$L$ the more pronounced is the rise of the large-$k$ tail of 
$E^p(k)$ (cf. the plots in  Fig. (\ref{fig:spect_L2}a) with red triangles 
and green asterisks for $L=100$ and $L=10$, respectively).

We can understand these trends qualitatively by noting
that, even at maximal extension, the size of a polymer is $\le \eta$ (the dissipation
scale). Thus, the polymers stretch at the expense of the fluid energy, which
cascades from the intermediate length scales to dissipative scales; this leads
to a reduction of $E(k)$ at the values of $k$ that correspond to these
intermediate scales. As the polymers relax, they feed energy to the fluid at
the deep-dissipation, i.e., large-$k$, scales; this leads to an enhancement in
the tail of $E(k)$ at large values of $k$. The reduction of energy in the
inverse-cascade, low-$k$ regime can be understood by noting that polymers
enhance the overall, effective viscosity of the fluid.  Indeed, in the limit 
$\tau_P \to 0$, $\nu \nabla^2 {\bf u} + \frac{\mu}{\tau_P} \nabla \cdot {f(r_P) C} 
\to (\nu+\mu) \nabla^2 {\bf u}$ ~\cite{bird87}.

\begin{figure*}
\includegraphics[width=0.325\linewidth]{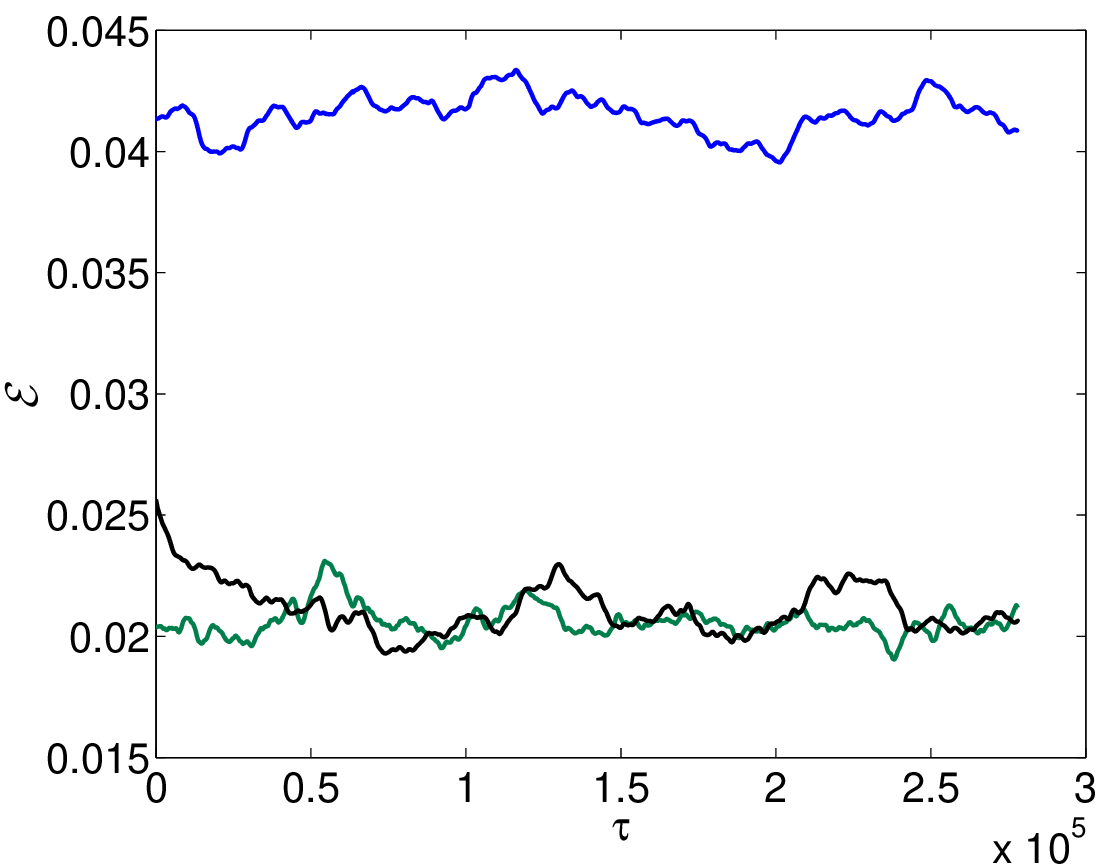}
\put(-45,80){ { {\large (a)} } }
\includegraphics[width=0.325\linewidth]{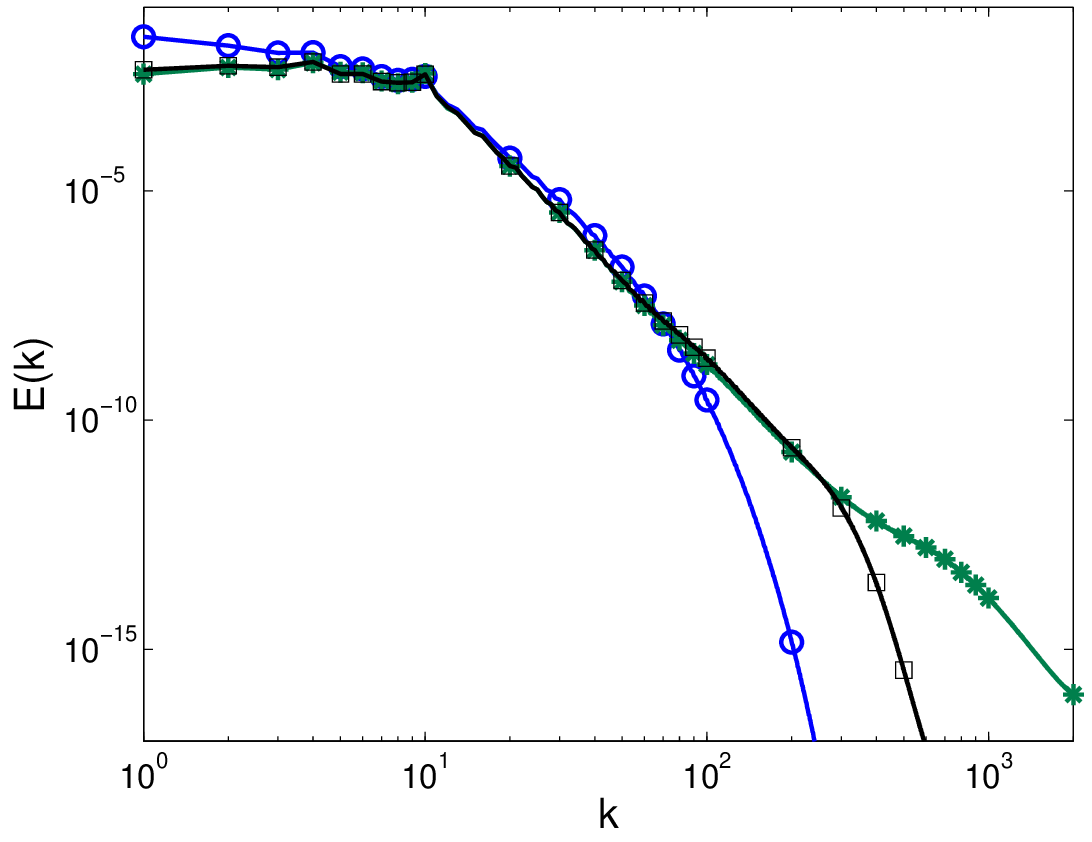}
\put(-45,100){ { {\large (b)} } }
\includegraphics[width=0.325\linewidth]{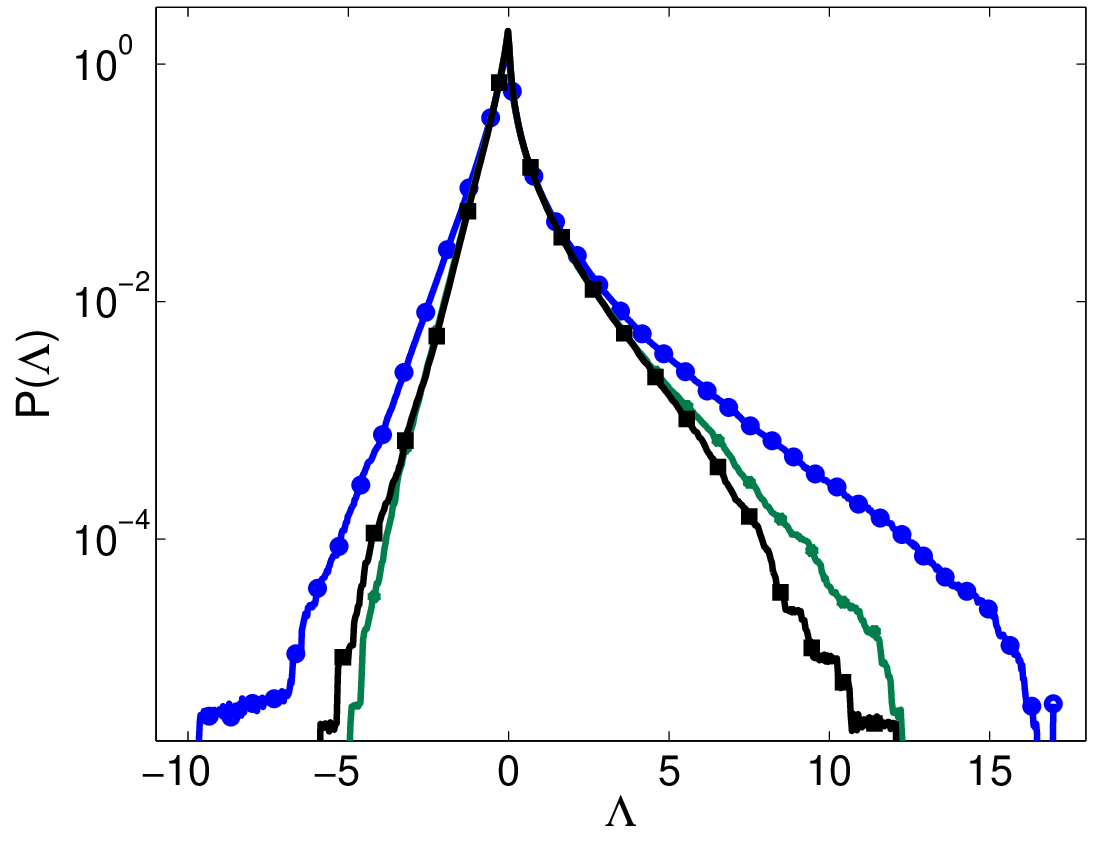}
\put(-45,100){ { {\large (c)} } }
\caption{\label{fig:scvis}\small(Color online) Plots with comparisons 
of the energy (a), energy spectra (b), and the PDF
of $\Lambda$ (c) from our DNS of the NS equations with the 
\textit{scale-dependent} viscosity $\nu_e(k)$ (black squares) and 
from the NS+FENE-P run $\tt R7$ (green stars). (We calculate 
$\nu_e(k)\equiv \nu + \Delta \nu(k)$ by substituting our data from 
run $\tt R7$ into Eq.~\eqref{eq:scvis}.) For reference, we also give plots 
of all these quantities for the NS equation with conventional, 
\textit{scale-independent} viscosity (blue circles).}
\end{figure*}

To understand quantitatively the effect of polymers on $E(k)$, in different regimes of $k$,
we must compare the fluid-energy spectra, with and without polymers (Fig.
(1b)).  This leads us naturally to define~\cite{procacciapoly,perlekar06,perlekar10} the effective, scale-dependent
viscosity $\nu_e(k)\equiv\nu+\Delta \nu(k)$, with 
\begin{eqnarray}
\Delta \nu(k)&\equiv&-\mu \sum_{k-1/2 < k' \le k+1/2} \frac{{\bf u}_{\bf k'} \cdot (\nabla \cdot {\cal J})_{\bf
-k'}}{[\tau_P{k}^2E^{p}(k)]}
\label{eq:scvis}
\end{eqnarray}
and $(\nabla \cdot {\cal J})_{\bf k}$ the Fourier transform of $\nabla \cdot
{\cal J}$. Figure (\ref{fig:eddr}c) shows that $\Delta \nu(k)>0$ for $k<30$,
where $E^{p}(k) < E^f(k)$, whereas, for large values of $k$, $\Delta \nu(k) <
0$, where $E^{p}(k) > E^f(k)$; the superscripts $f$ and $p$ stand,
respectively, for the fluid without and with polymers. 
To understand this dependence on $L$ we plot, in Fig.
(\ref{fig:spect_L2}b), the scale-dependent viscosity $\Delta \nu$
for these two representative values, namely, $L = 100$ (red triangles and run $\tt
R8$) and $L=10$ (green asterisks and run $\tt R9$). We find that
$\Delta \nu$ is positive and higher for $L = 100$, at small values
of $k$, than its counterpart for $L=10$; this explains why $E^p(k)$ is 
smaller for $L=100$ than for $L = 10$ at small $k$. For large values of $k$, 
$\Delta \nu$ is more negative for $L=100$ than for $L = 10$, so $E^p(k)$ is 
larger for $L=100$ than for $L = 10$. Note that $\Delta \nu (k)$ changes
its sign, from positive to negative, at a smaller value of $k$ for $L=100$ 
than for $L = 10$; therefore, the large-$k$ tail of $E^p(k)$ rises
above that of $E^f(k)$ at a smaller value of $k$ for $L = 100$ 
than for $L = 10$. By using $\nu_e(k)$,
which we obtain from our NS+FENE-P run $\tt R7$, we carry out a DNS of the NS
equation with $\nu$ replaced by $\nu_e(k)$; in Fig.~\ref{fig:scvis} we present
plots of the energy (left panel), energy spectra (middle panel), PDFs of
$\Lambda$ (right panel and Fig.~\ref{fig:lam_rP}), to compare the results of
this DNS with those of run $\tt R7$ (NS+FENE-P); the good agreement of these
results shows that the NS equation with the scale-dependent viscosity
$\nu_e(k)$ captures the essential effects of polymer additives on fluid
turbulence in run $\tt R7$ (NS+FENE-P).  The form of our effective viscosity
indicates that, at large length scales, in addition to the friction, polymers
also provide a dissipative mechanism. By contrast, at small length scales,
polymers inject energy back into the fluid.

Figure (\ref{fig:eddr}d) shows the suppression, by polymer additives, of
$\Pi(k) = \int_{k'}^{\infty} T(k')dk'$, where $T(k) = \int \hat{u_i}({\bf
-k})P_{ij}({\bf k}) \widehat{({\bf u} \times \boldsymbol \omega)_j}({\bf k}) {d
\Omega}$ and $P_{ij}({\bf k}) = \delta_{ij}-\frac {k_ik_j}{k^2} $.  The
suppression of the spectrum in the small-$k$ r\'egime, which has also been seen
in experiments ~\cite{amarouchene02} and low-resolution DNS (Fig. (4.12) of
Ref.~\cite{mussachiothesis}), signifies a reduction of the inverse cascade; the
enhancement of the spectrum in the large-$k$ r\'egime leads to the reduction in
$\Omega$ and ${\mathcal P}$ shown in Fig. (\ref{fig:eddr}a); to identify this
enhancement unambigouosly requires the run $\tt R10$, which is by far the
highest-resolution DNS of Eqs.  \eqref{ns}-\eqref{FENE} (with $256$ times more
collocation points than, say, Ref.~\cite{berti08}).

\begin{figure}
\hspace{-0.6cm}
\includegraphics[width=0.98\linewidth]{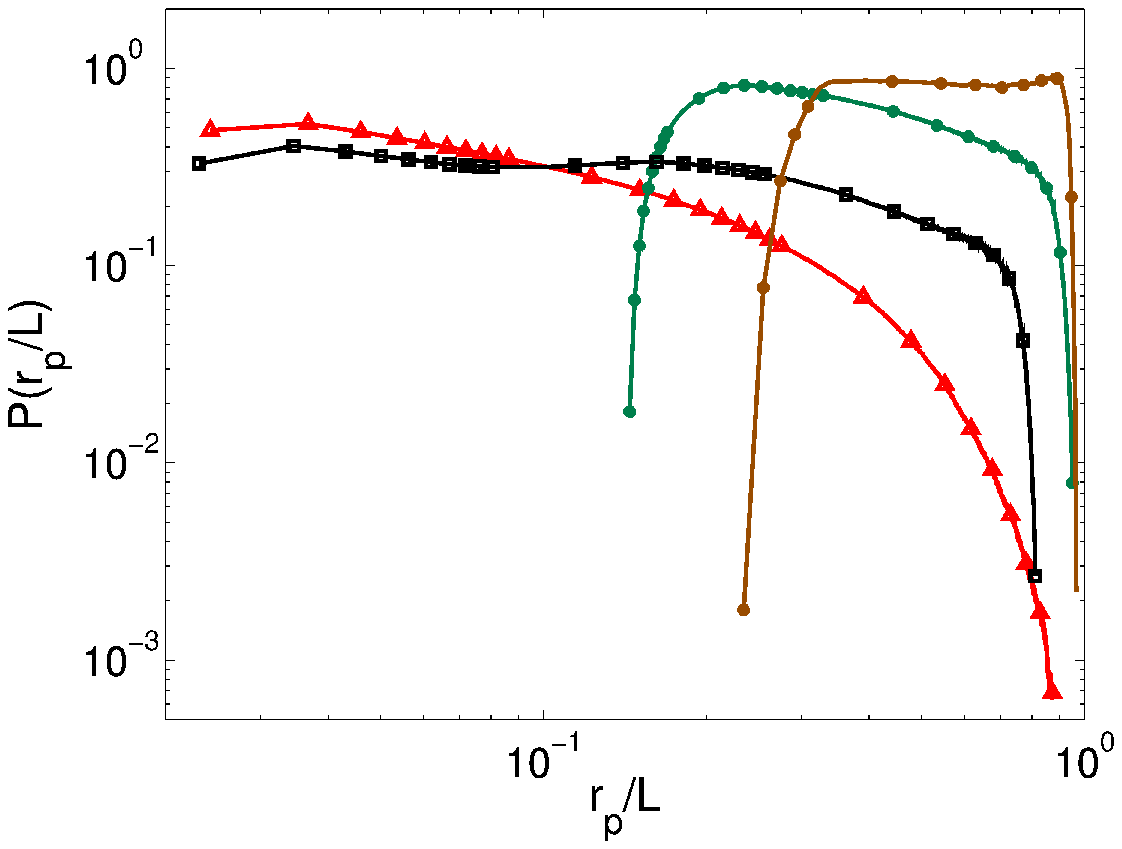}
\caption{\label{fig:PDF_rP}\small(Color online) 
PDFs of the scaled polymer extensions $P(r_P/L)$ versus $r_P/L$ for $c=0.1$ and $L=100$
(red triangles for run {\tt R8}), $c=0.4$ (black squares for run $\tt R8$), $c=0.1$ and $L
= 10$ (green asterisks for run $\tt R9 $), and  $c=0.1$ and $L = 6$ (brown dots for run
$\tt R1$).} 
\end{figure}
\begin{figure*}
\hspace{-0.5cm}
\includegraphics[width=0.325\linewidth]{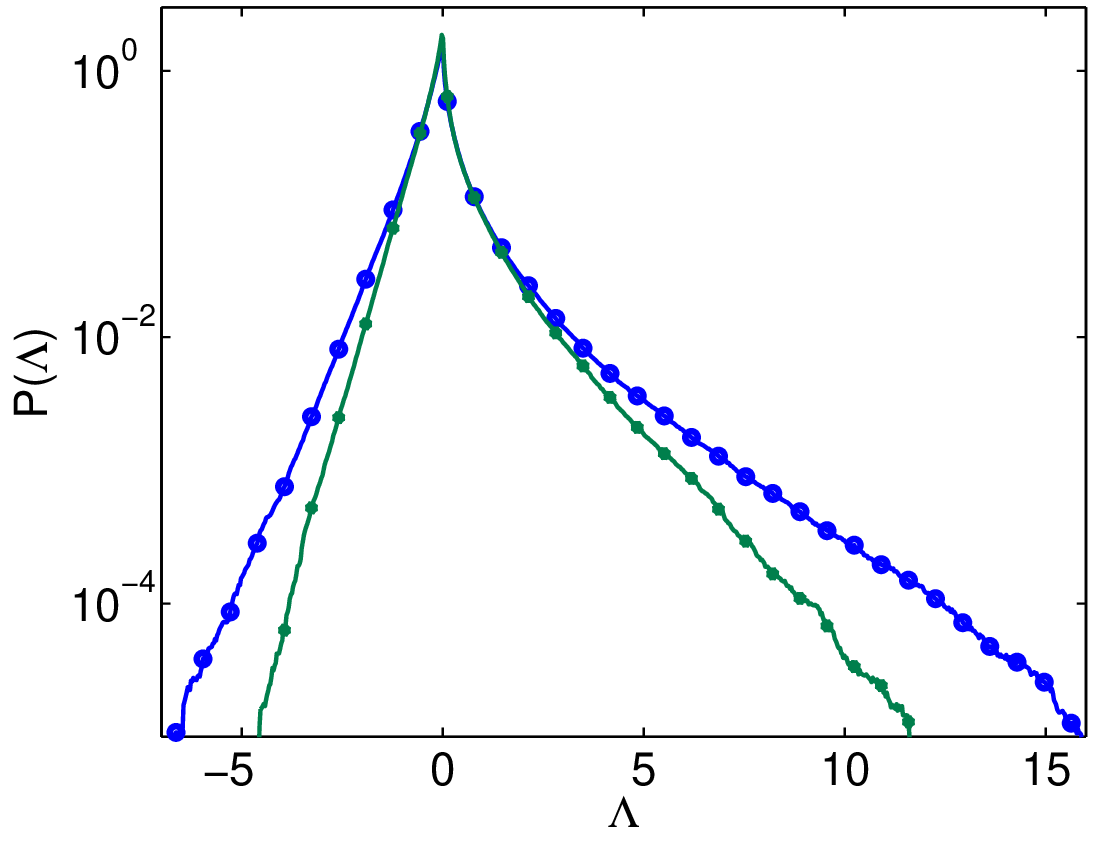}
\put(-40,105){ { {\large (a)} } }
\includegraphics[width=0.325\linewidth]{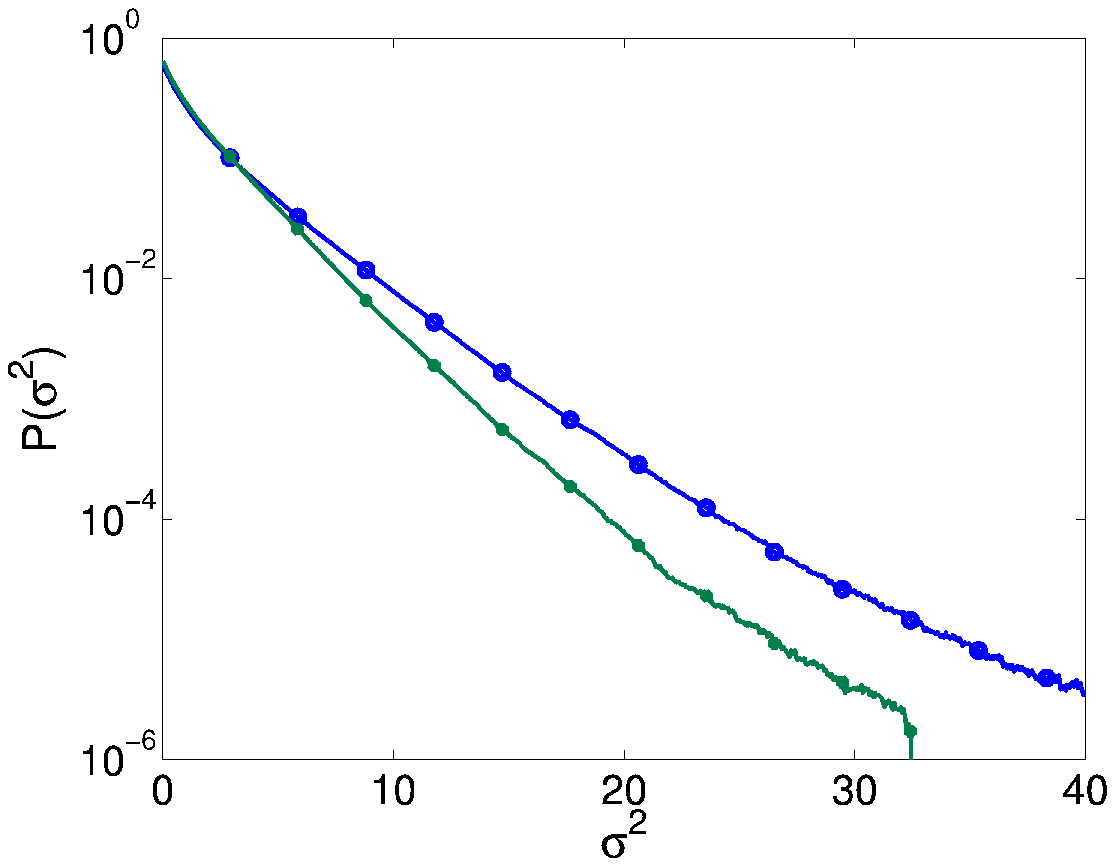}
\put(-40,105){ { {\large (b)} } }
\includegraphics[width=0.325\linewidth]{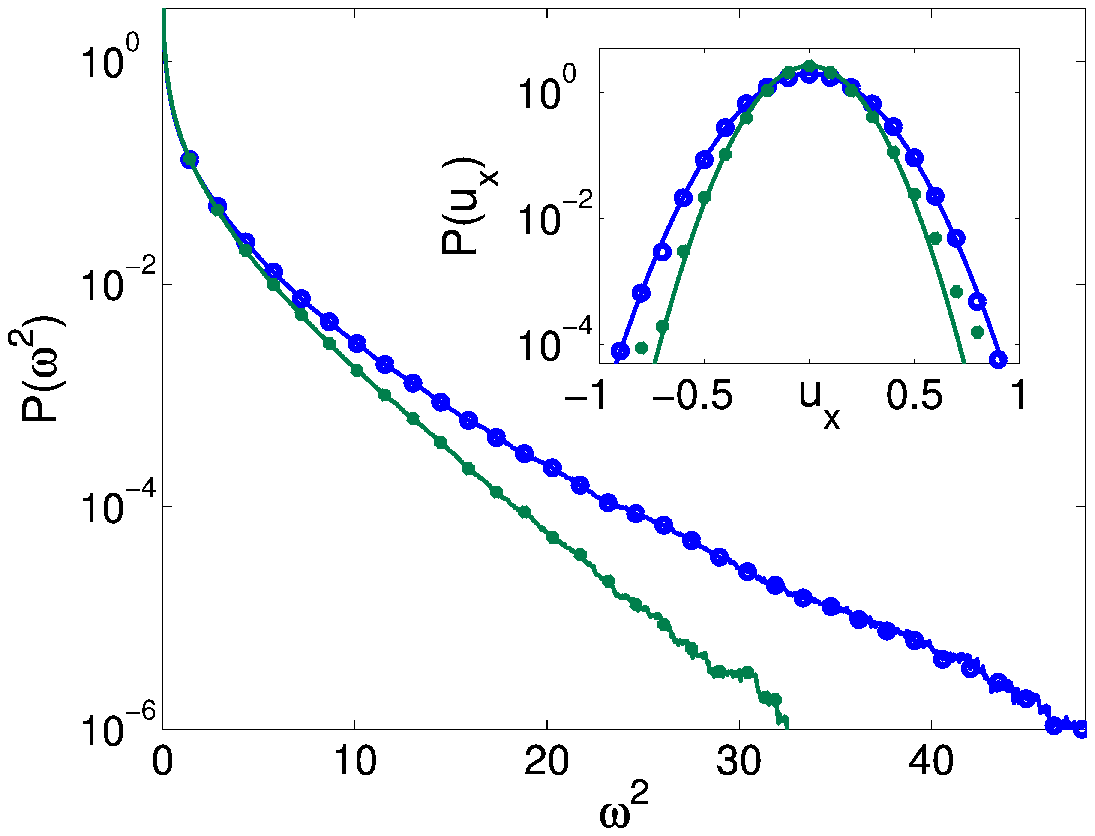}
\put(-125,105){ { {\large (c)} } }
\caption{\label{fig:str_vor}\small(Color online)
Probability distribution functions (PDFs) of (a) the Okubo-Weiss parameter $\Lambda$ for run $\tt R7$ ,
(b) $\sigma^2$ for run $\tt R7$, and 
(c) $\omega ^2$ [inset: a PDF of the velocity component $u_x$ for $c=0$ (blue circles for run $\tt R7$) with a fit
$(1/2) \exp(-u_x^2/12.5)$ (blue solid line), and for $c=0.2$ (green asterisks for run $\tt R7$) with a fit $(1/2.65) \exp(-u_x^2/20)$ (green solid line) (note that the addition
of polymers reduces the rms value of $u_x$)].
}
\end{figure*}
\begin{figure*}
\hspace{-0.5cm}
\includegraphics[width=0.47\linewidth]{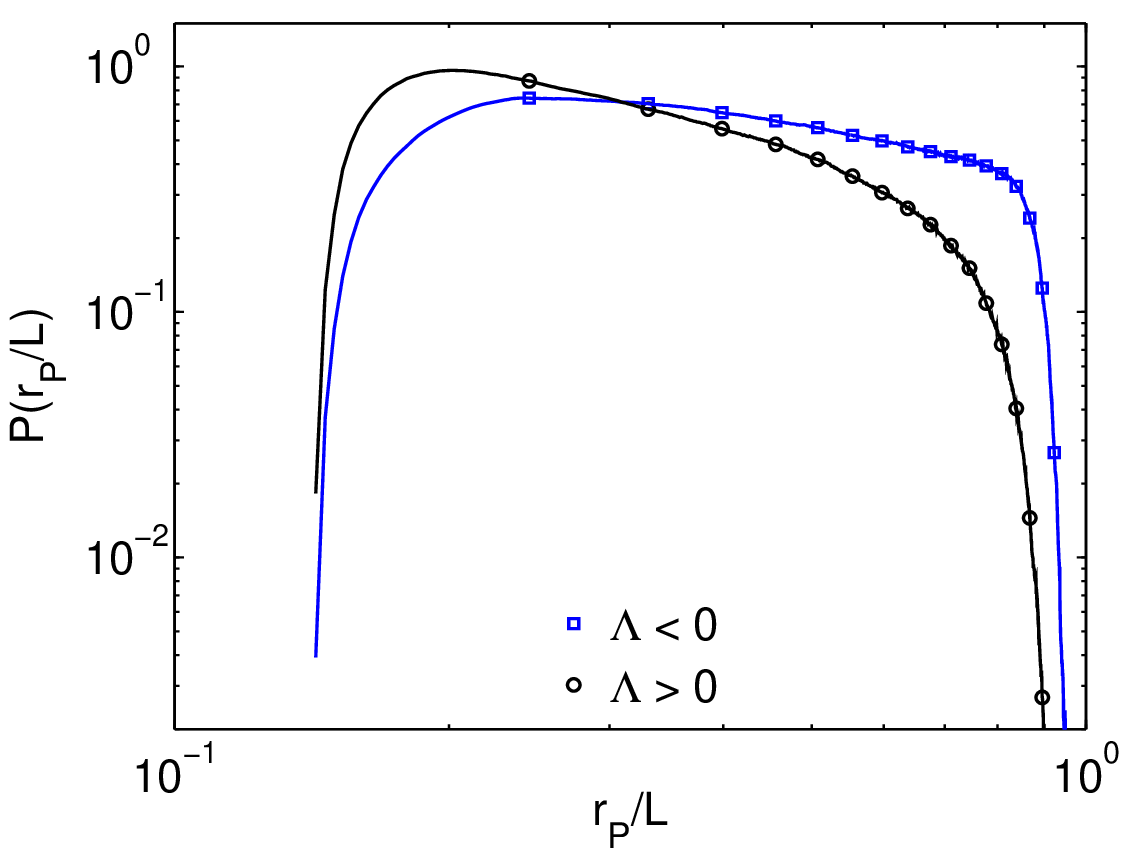}
\put(-195,160){ { {\large (a)} } }
\includegraphics[width=0.47\linewidth]{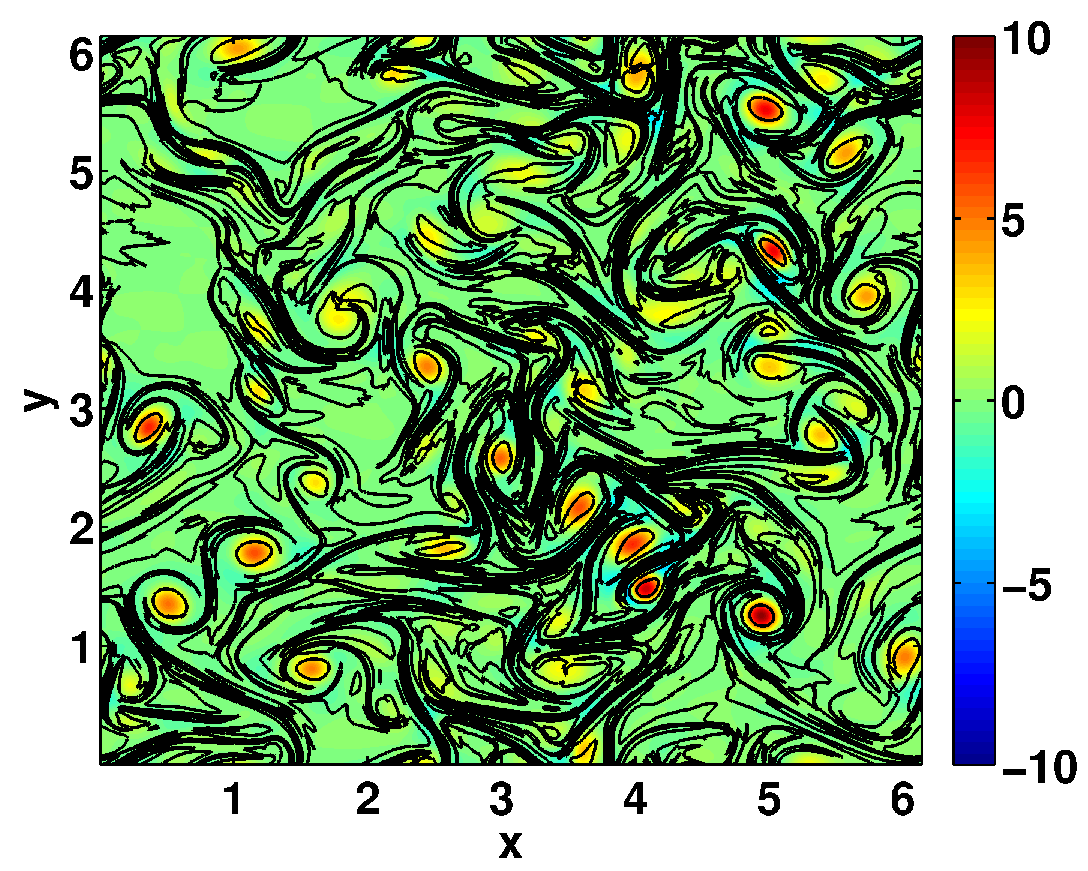}
\put(-195,160){ \large {\textcolor{magenta}{\large  (b)} } }
\caption{\label{fig:lam_rP}\small(Color online)
(a) Conditional PDF of ($r_P/L$) conditioned on $\Lambda$ for run $\tt R9$),
(b) a pseudocolor plot of $\Lambda$ superimposed on a contour plot of $r_P^2$ for run $\tt R10$.}
\end{figure*}

We now plot the PDF $P(r_P/L)$ versus $r_P/L$ in Fig. (\ref{fig:PDF_rP}) for
$c=0.1$ and $L=100$ (red triangles and run $\tt R8$), $c=0.4$ and $L=100$
(black squares and run $\tt R8$), and $c=0.1$ and $L=10$ (green asterisks and
run $\tt R9$).  The extension of the polymers is bounded between $\sqrt{2}\leq
r_P \leq L$. The lower bound, $r_P = \sqrt{2}$, corresponds to polymers in a
coiled state; near the upper bound, with $r_P\sim L$, the polymers are in a
stretched state. In Fig. (\ref{fig:PDF_rP}), we show that $P(r_P/L)$ shows a
distinct, power-law regime, with exponents that depends on $c, L,$ and $Wi$.
As $Wi$ increases, this exponent can go from a negative value to a positive
value, thus signalling a coil-stretch transition.

In Figs. (\ref{fig:str_vor}a), (\ref{fig:str_vor}b), and (\ref{fig:str_vor}c)
we present PDFs of $\Lambda$, $\sigma^2$, and $\omega^2$, respectively, for
$c=0$ (blue circles and run $\tt R7$) and $c=0.2$ (red triangles and run $\tt
R7$) to show that the addition of polymers suppresses large values of
$\Lambda$, $\sigma^2$, and $\omega^2$.  If we make scaled plots of PDFs such as
$P(\Lambda/\Lambda_{rms})$, then they fall on top of each other for different
values of $c$; this also holds for $P(\sigma^2/\sigma^2_{rms})$ and
$P(\omega^2/\omega^2_{rms})$.  The inset of Fig. (\ref{fig:str_vor}c) shows
that the PDF of any Cartesian component of {\bf u} is very close to a Gaussian.

The Fig.(\ref{fig:lam_rP}a) shows a conditional PDF of ($r_P/L$) conditioned on
$\Lambda$ for run $\tt R9$; this illustrates that polymers stretch
predominantly in strain-dominated regions; this is evident very strikingly in
Fig. (\ref{fig:lam_rP}b), which contains a superimposition of contours of
$r_P^2$ on a pseudocolor plot of $\Lambda$ (for a video sequence of such plots
see \cite{suppmat}).

\section{Conclusions}
\label{sec:Conclusions}

We have carried out the most extensive and high-resolution DNS of 2D,
homogeneous, isotropic fluid turbulence with polymer additives. We have used
the incompressible, 2D NS equation with air-drag-induced friction and polymer
additives; the latter have been modelled by using the
finitely-extensible-nonlinear-elastic-Peterlin (FENE-P) model for the
polymer-conformation tensor. We find that the inverse-cascade part of the
energy spectrum in 2D fluid turbulence is suppressed by the addition of
polymers. We demonstrate, for the first time, that the effect of polymers on
the forward-cascade part of the fluid energy spectrum in 2D is (a) a slight
reduction at intermediate wave numbers and (b) a significant enhancement in the
large-wave-number range, as in 3D; these features are resolved unambiguously by
our high-resolution DNS.  In addition, we find dissipation-reduction-type
phenomena~\cite{perlekar06,perlekar10}: polymers reduce the total fluid energy
and energy- and mean-square-vorticity- or enstrophy-dissipation rates.
However, as we have emphasized above, dissipation reduction is \textit{not the
only notable effect} of polymer additives; our extensive, high-resolution DNS
of 2D fluid turbulence with polymer additives yields good qualitative
agreement, in the low-$k$ r\'egime, with the fluid-energy spectra of
Ref.~\cite{amarouchene02}, and the $S_2(r)$ of Ref.~\cite{jun06}. In addition,
our study obtains new results and insights that will, we hope, stimulate new
experiments, which should be able to measure (a) the reduction of $\langle
\mathcal E (t) \rangle_t$, $\langle \Omega(t)\rangle_t$, and $\langle {\mathcal
P}(t)\rangle_t$ (Fig.(\ref{fig:eddr}a)), (b) the modification of $E^p(k)$ at
large $k$ (Fig.(\ref{fig:eddr}b)), (c) the $c$, $\tau_P$ and $L$ dependences of
$E^p(k)$ (Figs.(\ref{fig:spect}a),(\ref{fig:spect}b) and
(\ref{fig:spect_L2}a)), (d) the PDFs of $(r_P/L)$, $\Lambda$, $\sigma^2$, and
$\omega^2$, (e) the stretching of polymers in strain-dominated regions (Fig.
(\ref{fig:lam_rP}b)), and (f) the suppression of $F_6(r)$ at small $r$
(Fig.~(\ref{fig:s2fun})).

Two-dimensional fluid turbulence with polymer additives has been studied in
channel flows, both in experiments~\cite{kellayexpt} and via
DNS~\cite{kellaydns}; this DNS study uses the Oldroyd-B model, which does not
have a maximal-polymer-extension length and is, therefore, less realistic than
the FENE-P model we use.  These studies obtain energy spectra and second-order
structure functions that are qualitatively similar to those we obtain, except
at small length scales, which are not resolved in these channel-flow studies.
This shows, therefore, that energy spectra and structure functions, obtained
far away from walls, are not affected significantly by the walls.  Thus, our
studies are relevant to the bulk parts of wall-bounded flows too.  

\section{Acknowledgments}


We thank D. Mitra for discussions, CSIR, UGC, DST (India), and  the COST Action
MP006 for support, and SERC (IISc) for computational resources.  AG thanks the
grant from European Research Council under the European Community's Seventh
Framework Programme (FP7/2007-2013)/ERC Grant Agreement N. 279004.

\end{document}